\documentclass[graybox]{svmult}
\usepackage{amsfonts}
\usepackage{amsmath}
\usepackage{mathptmx}
\usepackage{helvet}
\usepackage{courier}
\usepackage{type1cm}
\usepackage{makeidx}
\usepackage{graphicx}
\usepackage{multicol}
\usepackage[bottom]{footmisc}
\usepackage{helvet}         
\usepackage{courier}        
\usepackage{type1cm}        
\usepackage{makeidx}         
\usepackage{graphicx}        
\usepackage{multicol}        
\usepackage[bottom]{footmisc}

\makeindex

\begin{document}

\title*{Self Sustained Traversable Phantom Wormholes and Gravity's Rainbow}
\author{Remo Garattini}

\titlerunning{Traversable Wormholes in Gravity's Rainbow}

\institute{Remo Garattini \at Universit\`{a} degli Studi di Bergamo, \\
Dipartimento di Ingegneria e scienze applicate,\\
Viale Marconi,5 24044 Dalmine (Bergamo) ITALY\\
I.N.F.N. - sezione di Milano, Milan, Italy, \email{remo.garattini@unibg.it}}
\maketitle

\abstract{We consider the effects of Gravity's Rainbow on the
self-sustained equation which is responsible to find new traversable
wormholes configurations powered by their own gravitational
quantum fluctuations. The form of the shape function considered is obtained by imposing the
equation of state $p_{r}=\omega \rho $. We investigate the size of the wormhole
as a function of the parameter $\omega$
in the phantom region. We discover that a wormhole which is traversable in principle,
 but not in practice, can be produced.}

\section{Introduction}

\label{sec:1}In recent years, several attempts to modify gravity at
Planckian or Transplanckian energy have been considered. The hope is to
include quantum gravitational effects in the description of physical
phenomena keeping under control the usual Ultraviolet (UV) divergences, avoiding
the usual regularization/renormalization scheme. To this purpose,
Noncommutative geometry, Gravity's Rainbow and Generalized Uncertainty
Principle (GUP) represent some possibilities to cure the divergences that
appear in general relativity\cite{RemoGRw,RGPN}. In particular, when Gravity's Rainbow is
considered, spacetime is endowed with two arbitrary functions $g_{1}\left(
E/E_{P}\right) $ and $g_{2}\left( E/E_{P}\right) $ having the following
properties%
\begin{equation}
\lim_{E/E_{P}\rightarrow 0}g_{1}\left( E/E_{P}\right) =1\qquad \text{and}%
\qquad \lim_{E/E_{P}\rightarrow 0}g_{2}\left( E/E_{P}\right) =1.
\label{g1g2}
\end{equation}%
$g_{1}\left( E/E_{P}\right) $ and $g_{2}\left( E/E_{P}\right) $ appear into
the solutions of the modified Einstein's Field Equations\cite{MagSmo}%
\begin{equation}
G_{\mu \nu }\left( E/E_{P}\right) =8\pi G\left( E/E_{P}\right) T_{\mu \nu
}\left( E/E_{P}\right) +g_{\mu \nu }\Lambda \left( E/E_{P}\right) ,
\end{equation}%
where $G\left( E/E_{P}\right) $ is an energy dependent Newton's constant,
defined so that $G\left( 0\right) $ is the low-energy Newton's constant and $%
\Lambda \left( E/E_{P}\right) $ is an energy dependent cosmological
constant. Usually $E$ is the energy associated to the particle deforming the
spacetime geometry. Since the scale of deformation involved is the Planck
scale, it is likely that spacetime itself begins to fluctuate in such a way
to produce a Zero Point Energy (ZPE). In absence of matter fields, the only
particle compatible with the deformed Einstein's gravity is the graviton.
What makes a ZPE calculation interesting is that it is strictly related to
the Casimir effect. Casimir effect has many applications and it can be
considered under different points of view, but it can also be used as a tool
to probe another appealing production of the gravitational field theory: a
wormhole. A wormhole is often termed Einstein-Rosen bridge because a
\textquotedblleft \textit{bridge}\textquotedblright\ connecting two
\textquotedblleft sheets\textquotedblright\ was the result obtained by A.
Einstein and N. Rosen in attempting to build a geometrical model of a
physical elementary "particle" that was everywhere finite and singularity
free\cite{ER}. It was J.A. Wheeler who introduced the term wormhole\cite{JAW}%
, although his wormholes were at the quantum scale. We have to wait for M.
S. Morris and K. S. Thorne\cite{MT} to see the subject of wormholes
seriously considered by the scientific community. To exist, traversable wormholes must violate
the null energy conditions, which means that the matter threading the
wormhole's throat has to be \textquotedblleft \textit{exotic}%
\textquotedblright . Usually classical matter satisfies the usual energy conditions,
while Casimir energy on a fixed background has the correct properties to
substitute the exotic matter: indeed, it is known that, for different
physical systems, Casimir energy is negative. Usually one considers some
matter or gauge fields which contribute to the Casimir energy necessary to
the traversability of the wormholes, nevertheless nothing forbids to use the
Casimir energy of the graviton on a background of a traversable wormhole. In
this way, one can think that the quantum fluctuations of the gravitational
field of a traversable wormhole are the same ones which are responsible to
sustain traversability. Note that in Ref.\cite{DBGL}%
, the ZPE was used as an indicator for a topology change without a Gravity's
Rainbow scheme, while in Ref.\cite{RGFL}, it has been shown that a topology change is a ZPE consequence
induced by Gravity's Rainbow. In this contribution we would like to probe ZPE with the help of Gravity's Rainbow
 and an equation of state.

\section{Gravity's Rainbow and the Equation of State}

\label{p1}In Schwarzschild coordinates, the traversable wormhole metric can
be cast into the form%
\begin{equation}
ds^{2}=-\exp \left( -2\phi \left( r\right) \right) dt^{2}+\frac{dr^{2}}{1-%
\frac{b\left( r\right) }{r}}+r^{2}d\Omega ^{2}.  \label{metric}
\end{equation}%
where $\phi \left( r\right) $ is called the redshift function, while $%
b\left( r\right) $ is called the shape function and where $d\Omega
^{2}=d\theta ^{2}+\sin ^{2}\theta d\phi ^{2}$ is the line element of the
unit sphere. Using the Einstein field equation%
\begin{equation}
G_{\mu \nu }=8\pi GT_{\mu \nu },
\end{equation}%
in an orthonormal reference frame, we obtain the following set of equations%
\begin{equation}
\rho \left( r\right) =\frac{1}{8\pi G}\frac{b^{\prime }}{r^{2}},
\label{rhob}
\end{equation}%
\begin{equation}
p_{r}\left( r\right) =\frac{1}{8\pi G}\left[ \frac{2}{r}\left( 1-\frac{%
b\left( r\right) }{r}\right) \phi ^{\prime }-\frac{b}{r^{3}}\right] ,
\label{pr}
\end{equation}%
\begin{equation}
p_{t}\left( r\right) =\frac{1}{8\pi G}\left( 1-\frac{b\left( r\right) }{r}%
\right) \left[ \phi ^{\prime \prime }+\phi ^{\prime }\left( \phi ^{\prime }+%
\frac{1}{r}\right) \right] -\frac{b^{\prime }r-b}{2r^{2}}\left( \phi
^{\prime }+\frac{1}{r}\right) ,  \label{pt}
\end{equation}%
in which $\rho \left( r\right) $ is the energy density, $p_{r}\left(
r\right) $ is the radial pressure, and $p_{t}\left( r\right) $ is the
lateral pressure. Using the conservation of the stress-energy tensor, in the
same orthonormal reference frame, we get%
\begin{equation}
p_{r}^{\prime }=\frac{2}{r}\left( p_{t}-p_{r}\right) -\left( \rho
+p_{r}\right) \phi ^{\prime }.
\end{equation}%
When Gravity's Rainbow comes into play, the line element $\left( \ref{metric}%
\right) $ becomes\cite{MagSmo} 
\begin{equation}
ds^{2}=-\exp \left( -2\phi \left( r\right) \right) \frac{dt^{2}}{%
g_{1}^{2}\left( E/E_{P}\right) }+\frac{dr^{2}}{\left( 1-\frac{b\left(
r\right) }{r}\right) g_{2}^{2}\left( E/E_{P}\right) }+\frac{r^{2}}{%
g_{2}^{2}\left( E/E_{P}\right) }d\Omega ^{2}\,  \label{dS}
\end{equation}%
and the Einstein's Field Equations $\left( \ref{rhob}\right) $, $\left( \ref{pr}%
\right) $ and $\left( \ref{pt}\right) $ can be rearranged to give%
\begin{equation}
b^{\prime }=\frac{8\pi G\rho \left( r\right) r^{2}}{g_{2}^{2}\left(
E/E_{P}\right) },  \label{b'g2}
\end{equation}%
\begin{equation}
\phi ^{\prime }=\frac{b+8\pi Gp_{r}r^{3}/g_{2}^{2}\left( E/E_{P}\right) }{%
2r^{2}\left( 1-\frac{b\left( r\right) }{r}\right) }.  \label{phig2}
\end{equation}%
Now, we introduce the equation of state $p_{r}=\omega \rho $, and using Eq.$%
\left( \ref{b'g2}\right) $, Eq.$\left( \ref{phig2}\right) $ becomes%
\begin{eqnarray}
\phi ^{\prime } &=&\frac{b+8\pi G\left( \omega g_{2}^{2}\left(
E/E_{P}\right) b^{\prime }\left( r\right) /\left( 8\pi Gr^{2}\right) \right)
r^{3}/g_{2}^{2}\left( E/E_{P}\right) }{2r^{2}\left( 1-\frac{b\left( r\right) 
}{r}\right) }  \notag \\
&=&\frac{b+\omega b^{\prime }r}{2r^{2}\left( 1-\frac{b\left( r\right) }{r}%
\right) }.
\end{eqnarray}%
It is immediate to see that the equation related the redshift is unchanged
and can be set to a constant with respect to the radial distance if%
\begin{equation}
b+\omega b^{\prime }r=0.  \label{bb'}
\end{equation}%
The integration of this simple equation leads to%
\begin{equation}
b\left( r\right) =r_{0}\left( \frac{r_{0}}{r}\right) ^{\frac{1}{\omega }},
\label{shape}
\end{equation}%
where we have used the condition $b\left( r_{t}\right) =r_{t}$. In this
situation, the line element $\left( \ref{dS}\right) $ becomes%
\begin{equation}
ds^{2}=-\frac{A}{g_{1}^{2}\left( E/E_{P}\right) }dt^{2}+\frac{dr^{2}}{%
1-\left( \frac{r_{0}}{r}\right) ^{1+\frac{1}{\omega }}g_{2}^{2}\left(
E/E_{P}\right) }+\frac{r^{2}}{g_{2}^{2}\left( E/E_{P}\right) }d\Omega ^{2},
\label{line}
\end{equation}%
where $A$ is a constant coming from $\phi ^{\prime }=0$ which can be set to
one without loss of generality. The parameter $\omega $ is restricted by the
following conditions%
\begin{equation}
b^{\prime }\left( r_{0}\right) <1;\qquad \frac{b\left( r\right) }{r}\underset%
{r\rightarrow +\infty }{\rightarrow 0}\qquad \Longrightarrow \qquad \left\{ 
\begin{array}{c}
\omega >0 \\ 
\omega <-1%
\end{array}%
\right. .
\end{equation}

\section{Self-sustained Traversable Wormholes, Gravity's Rainbow and Phantom
Energy}

\label{p2}In this Section we shall consider the formalism outlined in detail
in Refs. \cite{Remo,Remo1}, where the graviton one loop contribution to a
classical energy in a wormhole background is used. A traversable wormhole is
said to be \textquotedblleft \textit{self sustained}\textquotedblright\ if%
\begin{equation}
H_{\Sigma }^{(0)}=-E^{TT},  \label{SS}
\end{equation}%
where $E^{TT}$ is the total regularized graviton one loop energy and $%
H_{\Sigma }^{(0)}$ is the classical term. When we deal with spherically
symmetric line element, the classical Hamiltonian reduces to%
\begin{align}
H_{\Sigma }^{(0)}& =\int_{\Sigma }\,d^{3}x\left[ \,\left( 16\pi G\right)
G_{ijkl}\pi ^{ij}\pi ^{kl}-\frac{\sqrt{g}}{16\pi G}\!{}\!\,\ R\right]  \notag
\\
& =-\frac{1}{16\pi G}\int_{\Sigma }\,d^{3}x\,\sqrt{g}\,R\,=-\frac{1}{2G}%
\int_{r_{0}}^{\infty }\,\frac{dr\,r^{2}}{\sqrt{1-b(r)/r}}\,\frac{b^{\prime
}(r)}{r^{2}g_{2}\left( E/E_{P}\right) }\,,
\end{align}%
where we have used the explicit expression of the scalar curvature in three
dimensions in terms of the shape function. $G_{ijkl}$ is the super-metric
and $\pi ^{ij}$ is the super-momentum. Note that, in this context, the
kinetic term disappears. Note also that boundary terms become important when
one compares different configurations like Wormholes and Dark Stars\cite%
{DBGL} or Wormholes and Gravastars\cite{JHEP}. With the help of Eq.$\left( %
\ref{bb'}\right) $, the classical energy becomes%
\begin{equation}
H_{\Sigma }^{(0)}=\frac{1}{2G}\int_{r_{0}}^{\infty }\,\frac{dr\,r^{2}}{\sqrt{%
1-b(r)/r}}\,\frac{b(r)}{r^{3}g_{2}\left( E/E_{P}\right) \omega }
\end{equation}%
and following Ref.\cite{RGFSNL}, the self-sustained equation $\left( \ref{SS}%
\right) $ becomes 
\begin{equation}
-\frac{b(r)}{2Gr^{3}g_{2}\left( E/E_{P}\right) \omega }=\frac{2}{3\pi ^{2}}%
\left( I_{1}+I_{2}\right) \,,  \label{ETT}
\end{equation}%
where the r.h.s. of Eq.$\left( \ref{ETT}\right) $ is represented by%
\begin{equation}
I_{1}=\int_{E^{\ast }}^{\infty }E\frac{g_{1}\left( E/E_{P}\right) }{%
g_{2}^{2}\left( E/E_{P}\right) }\frac{d}{dE}\left( \frac{E^{2}}{%
g_{2}^{2}\left( E/E_{P}\right) }-m_{1}^{2}\left( r\right) \right) ^{\frac{3}{%
2}}dE\,  \label{I1}
\end{equation}%
and%
\begin{equation}
I_{2}=\int_{E^{\ast }}^{\infty }E\frac{g_{1}\left( E/E_{P}\right) }{%
g_{2}^{2}\left( E/E_{P}\right) }\frac{d}{dE}\left( \frac{E^{2}}{%
g_{2}^{2}\left( E/E_{P}\right) }-m_{2}^{2}\left( r\right) \right) ^{\frac{3}{%
2}}dE\,,  \label{I2}
\end{equation}%
respectively. $E^{\ast }$ is the value which annihilates the argument of the
root while $m_{1}^{2}\left( r\right) $ and $m_{2}^{2}\left( r\right) $ are
two r-dependent effective masses. Of course, $I_{1}$ and $I_{2}$ are finite
for appropriate choices of the Rainbow's functions $g_{1}\left(
E/E_{P}\right) $ and $g_{2}\left( E/E_{P}\right) $. With the help of the
EoS, one finds%
\begin{equation}
\left\{ 
\begin{array}{c}
m_{1}^{2}\left( r\right) =\frac{6}{r^{2}}\left( 1-\frac{b\left( r\right) }{r}%
\right) +\frac{3}{2r^{3}\omega }b\left( r\right) \left( \omega +1\right) \\ 
\\ 
m_{2}^{2}\left( r\right) =\frac{6}{r^{2}}\left( 1-\frac{b\left( r\right) }{r}%
\right) +\frac{3}{2r^{3}\omega }b\left( r\right) \left( \frac{1}{3}-\omega
\right)%
\end{array}%
\right.
\end{equation}%
and on the throat, the effective masses reduce to%
\begin{equation}
\begin{tabular}{c}
$m_{1}^{2}\left( r_{0}\right) =\frac{3}{2r_{0}^{2}\omega }\left( \omega
+1\right) \qquad \left\{ 
\begin{array}{l}
>0\qquad \mathrm{when\qquad }\omega >0\qquad \mathrm{or\qquad }\omega <-1 \\ 
<0\qquad \mathrm{when\qquad }-1<\omega <0%
\end{array}%
\right. $ \\ 
\\ 
$m_{2}^{2}\left( r_{0}\right) =\frac{3}{2r_{0}^{2}\omega }\left( \frac{1}{3}%
-\omega \right) \qquad \left\{ 
\begin{array}{l}
>0\qquad \mathrm{when\qquad }1/3>\omega >0 \\ 
<0\qquad \mathrm{when\qquad }\omega >1/3\qquad \mathrm{or\qquad }\omega <0%
\end{array}%
\right. $%
\end{tabular}%
.  \label{m1m2}
\end{equation}%
However, to have values of $\omega $ compatible with the traversability
condition, only the cases with $\omega >0$ and $\omega <-1$ are allowed. It
is easy to see that if we assume%
\begin{equation}
g_{1}\left( E/E_{P}\right) =1\qquad g_{2}\left( E/E_{P}\right) =\left\{ 
\begin{array}{c}
1\qquad \mathrm{when}\qquad E<E_{P} \\ 
\\ 
E/E_{P}\qquad \mathrm{when}\qquad E>E_{P}\qquad%
\end{array}%
\right. \,,  \label{rel}
\end{equation}%
Eq.$\left( \ref{ETT}\right) $ becomes, close to the throat,%
\begin{equation}
-\frac{1}{2Gr_{0}^{2}\omega }=\frac{2}{\pi ^{2}}\left( \int_{\sqrt{%
m_{1}^{2}\left( r\right) }}^{E_{P}}E^{2}\sqrt{E^{2}-m_{1}^{2}\left(
r_{0}\right) }dE\,+\int_{\sqrt{m_{2}^{2}\left( r\right) }}^{E_{P}}E^{2}\sqrt{%
E^{2}-m_{2}^{2}\left( r_{0}\right) }dE\right) \,,  \label{ETT1}
\end{equation}%
where $m_{1}^{2}\left( r_{0}\right) $ and $m_{2}^{2}\left( r_{0}\right) $
have been defined in $\left( \ref{m1m2}\right) $. Since the r.h.s. is
certainly positive, in order to have real solutions compatible with
asymptotic flatness, we need to impose $\omega <-1$, that it means that we
are in the Phantom regime. With this choice, the effective
masses $\left( \ref{m1m2}\right) $ become, on the throat%
\begin{equation}
m_{1}^{2}\left( r_{0}\right) =\frac{3}{2r_{0}^{2}\omega }\left( \omega
+1\right) \qquad m_{2}^{2}\left( r_{0}\right) =-\frac{3}{2r_{0}^{2}\omega }%
\left( \frac{1}{3}-\omega \right)
\end{equation}%
and Eq.$\left( \ref{ETT1}\right) $ simplifies into%
\begin{equation}
1=-\frac{4r_{0}^{2}\omega }{\pi ^{2}E_{P}^{2}}\left( \int_{\sqrt{%
m_{1}^{2}\left( r_{0}\right) }}^{E_{P}}E^{2}\sqrt{E^{2}-\frac{3}{%
2r_{0}^{2}\omega }\left( \omega +1\right) }dE+\int_{0}^{E_{P}}E^{2}\sqrt{%
E^{2}+\frac{3}{2r_{0}^{2}}\left\vert \frac{1}{3\omega }-1\right\vert }%
dE\right) \,  \label{ETT2}
\end{equation}%
The solution can be easily computed numerically and we find%
\begin{align}
-1& \geq \omega \geq -4.5  \notag \\
2.038& \geq x\geq 1.083.
\end{align}
Therefore we can conclude that a wormhole which is traversable in principle, but not in practice, can be produced joining
Gravity's Rainbow and phantom energy. Of course, the result is strongly dependent on the rainbow's functions which, nonetheless must be chosen in such a way to give finite results for the one loop integrals $\left( \ref{I1}\right) $ and $\left( \ref{I2}\right) $.

\end{document}